\documentclass[manuscript]{aastex62}

\usepackage{amssymb,amsmath}
\usepackage{bm}
\usepackage{ulem}
\usepackage{comment}
\usepackage{color}

\received{2019 August 12}
\revised{2019 September 27}
\accepted{2019 September 27}

\shorttitle{In-Situ Formation of the Uranian Satellites}
\shortauthors{Ishizawa et al.}

\begin{document}

\title{Can the Uranian Satellites Form from a Debris Disk Generated by a Giant Impact?}

\correspondingauthor{Yuya Ishizawa}
\email{ishizawa@kusastro.kyoto-u.ac.jp}

\author{Yuya Ishizawa}
\affiliation{Department of Astronomy, Kyoto University,  Kitashirakawa-Oiwake-cho, Sakyo-ku, Kyoto 606-8502, Japan}

\author[0000-0003-1242-7290]{Takanori Sasaki}
\affiliation{Department of Astronomy, Kyoto University,  Kitashirakawa-Oiwake-cho, Sakyo-ku, Kyoto 606-8502, Japan}

\author[0000-0002-6638-7223]{Natsuki Hosono}
\affiliation{Japan Agency for Marine-Earth Science and Technology, 2-15, Natsushima-cho, Yokosuka-city, Kanagawa 237-0061, Japan}
\affiliation{RIKEN Center for Computational Science, 7-1-26 Minatojima-minami-machi, Chuo-ku, Kobe, Hyogo, Japan}

\begin{abstract}

Hydrodynamic simulations of a giant impact to proto-Uranus indicated that such an impact could tilt its rotational axis and produce a circumplanetary debris disk beyond the corotation radius of Uranus.
However, whether Uranian satellites can actually be formed from such a wide disk remains unclear.
Herein, we modeled a wide debris disk of solids with several initial conditions inferred from the hydrodynamic simulations, and performed $N$-body simulations to investigate in-situ satellite formation from the debris disk. 
We also took account of orbital evolutions of satellites due to the planetary tides after the growth of satellites. 
We found that, in any case, the orbital distribution of the five major satellites could not be reproduced from the disk as long as the power index of its surface density is similar to that of the disk generated just after the giant impact: Satellites in the middle region obtained much larger masses than Ariel or Umbriel, while the outermost satellite did not grow to the mass of Oberon. 
Our results indicate that we should consider the thermal and viscous evolution of the evaporated disk after the giant impact to form the five major satellites through the in-situ formation scenario.
On the other hand, the small inner satellites would be formed from the rings produced by the disrupted satellites which migrated from around the corotation radius of Uranus due to the planetary tides.

\end{abstract}


\section{Introduction} \label{sec:intro}

The Uranian mass, $M_{\rm U}$, is 8.68$\times 10^{25}$kg and is roughly 14.5 times that of Earth.
The axial tilt of Uranus is around 98$^\circ$ and is very large compared to those of other planets in the Solar System.
The origin  of this large axial tilt remains unclear. One hypothesis is that during the formation of the Solar System, an Earth-sized protoplanet  collided with Uranus and tilted its rotational axis \citep{Saf66}.

Uranus has 27 satellites that  are divided into three groups: 18 regular satellites, including 13 inner satellites and five major satellites, and nine irregular satellites.
The five major satellites account for more than 99$\%$ of the total mass of all Uranian satellites, $M_{\rm tot}$, where $M_{\rm tot} \simeq 1.05\times10^{-4}M_{\rm U}$.
 All regular Uranian satellites, except for Miranda, have very small  eccentricities ($\leq 0.01$) and small orbital inclinations ($\leq 0.5$).
Uranus also has rings with negligible inclinations ($\leq 0.1$).
These features mean that Uranus' regular satellites and rings lie almost on its equatorial plane.
The Uranian system, except for the irregular satellites, is tilted as a whole.

Several scenarios have been proposed to explain the origins of regular satellites around gas/ice giants, including the Uranian satellites.

The relatively large regular satellites orbiting around giant planets may have formed in circumplanetary disks at the end stage of the formation of the Solar System \citep{CW06,Sas10}.
\citet{CW06} modeled satellite growth in an actively supplied circumplanetary disk sustained by a time-dependent inflow of gas and solids from the solar nebula.
A satellite grew in such a disk until it fell onto the planet because of orbital decay due to the gravitational interaction with gas in the disk (e.g., \citealp{War86}). Therefore, the maximum satellite mass is determined by a balance between the timescale of satellite growth and the orbital decay of the satellite.
They showed that the ratio of the total satellite mass to the host planet's mass is commonly regulated to $\sim 10^{-4}$; the mass fraction of the Uranian satellite system is also roughly $\sim 10^{-4}$. However, their model required additional explanations for the large axial tilt of Uranus, such as secular perturbations by a temporally captured satellite over a long term \citep{BL10}.

\citet{CC12} proposed  an analytical model for the accretion and orbital evolution of satellites from a disk of solid materials around a planet.
The inner edge of the disk is the planet's radius and the outer edge is the Roche limit, within which the planet's tidal forces prevent the aggregation of the solid materials.
Such a tidal disk spreads owing to  the disk's viscosity \citep{Dai01} beyond the Roche limit, and then, a satellite forms via the disk's mass flow beyond this limit.
In their model, a satellite that grows outside the Roche limit  migrates outward by receiving a positive tidal torque from the planet and the disk.
Similarly, a new satellite forms outside the edge of the disk and migrates.
As satellites form, the disk's mass and mass flow both decrease, and therefore, the mass of formed satellites gradually decreases with generations.
A satellite's migration speed decreases with its orbital radius, and an inner satellite can reach the region where an outer satellite can merge with it.
When the differences between the masses and orbital radii of satellites side by side  become large enough, the satellites do not merge because  their migration speed decreases with the planet's mass and the timescale of satellite accretion increases as the disk mass decreases.
They analytically investigated the accretion and migration of satellites in terms of several parameters. 
They note that their model can explain the mass distribution with orbital radii of regular satellites of Saturn, Uranus, and Neptune.
Their model assumes that formed satellites do not perturb the mutual orbit and disk.
It also assumes significantly strong tidal dissipation inside Saturn, given by the tidal quality factor $Q_{\rm p} = 1680$, and they applied this value to the other planets (see Section \ref{sc:dis}).

\citet{Hyo15} performed direct numerical simulations of satellite formation in the disks modeled by \citet{CC12} to investigate a more realistic dynamical effect on the accretion and orbital evolution of satellites.
They confirmed that at least 1--2 satellites are formed from the disks.
The origin of a tidal disk is not explicit; however, they suggest that a tidal disk can be produced by tidal disruption of a passing heliocentric comet  or a satellite falling inside the Roche limit.
This model also needed an additional scenario for the origin of Uranus' large axial tilt.

\begin{figure}[htb]
  \centering
  \plotone{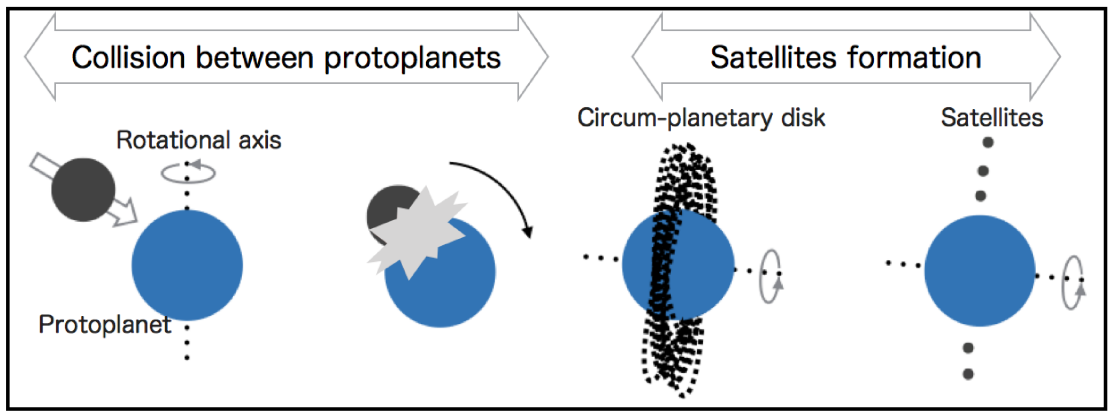}
  \caption{Schematic of satellite formation in giant impact scenario. First, two protoplanets collide with each other and the materials of these two bodies are ejected. Second, satellites form from the circumplanetary disk of ejected materials.}
\label{fg:gi}
\end{figure}

Another proposed scenario for Uranian satellite formation is the giant impact scenario.
A giant impact is a high energetic collision between protoplanets during the end stages of the formation of the Solar System.
In this scenario, one or more satellites can form from a circumplanetary disk generated by a large planetary body's impact with a protoplanet.
Figure \ref{fg:gi} shows a schematic of satellite formation in the giant impact scenario.
This scenario can simultaneously explain both the large axial tilt and the formation of the regular satellites of Uranus.

A giant impact is usually modeled using the smoothed particle hydrodynamics (SPH) method, in which fluid elements are represented by particles. 
\citet{Sla92} performed SPH simulations of collisions between proto-Uranus and an impactor with its mass of 1--3 Earth masses. 
In their simulations, the system's total angular momentum was considered a variable parameter.
They concluded that a fairly large parameter range of giant impacts could produce Uranus' present rotational period and large axial tilt.
However, the numerical resolutions of their SPH simulations are not high enough to determine  whether Uranian regular satellites could form by accreting materials ejected in orbit by the giant impact.

Recently, \citet{Keg18} performed SPH simulations describing a giant impact into proto-Uranus with much higher resolution compared to \citet{Sla92}.
They suggested that a large amount of rock/ice materials would be ejected beyond the corotation radius of Uranus by a high energetic impact. The giant impact can produce a disk with enough material for regular satellites to form. 
However, until recently, the formation of Uranian satellites via the giant impact scenario had not been investigated numerically.

The present study adopts the giant impact scenario as a possible process for the formation of Uranian regular satellites.
We model a wide disk around Uranus and investigate the in-situ formation  of the Uranian regular satellites using $N$-body simulations.

\section{Calculation method}

We considered a model in which satellites grow within a wide circumplanetary disk of solids and performed $N$-body simulations to investigate the in-situ formation of Uranian regular satellites.
An $N$-body simulation describes a dynamical system of particles, mainly under gravity.
Here, a particle represents a small rock-ice solid body that eventually forms a satellite.
The calculations consider gravitational interaction, collision, and  merger between particles.
In the following $N$-body simulation, mass, distance, and time are respectively normalized by the Uranian mass $M_{\rm U}$, Roche limit $a_{\rm R}$ given by
\begin{eqnarray}
 a_{\rm R} = 2.456 \left(\frac{\rho}{\rho_{\rm U}}\right)^{-\frac{1}{3}}R_{\rm U} = 2.38R_{\rm U}, 
\end{eqnarray}
and inverse of angular velocity at Roche limit $\Omega_{\rm R}^{-1}$ given by
\begin{eqnarray}
 \Omega_{\rm R} = \sqrt{\frac{GM_{\rm U}}{a_{\rm R}^3}},
\end{eqnarray}
where $G$ is the gravitational constant; $\rho=1.40$ g cm$^{-3}$, the mean density of the satellite system; $\rho_{\rm U}=1.27$ g cm$^{-3}$, the density of Uranus; and $R_{\rm U} = 0.42 a_{\rm R}$, the Uranian radius.
Inside the Roche limit, the planet's tidal force exceeds the relatively small body's self-gravity.
Therefore, a satellite cannot accrete inside the Roche limit; however, it can do so outside this limit.

\subsection{Numerical method}

Particle orbits are calculated according to the following equation of motion.
\begin{eqnarray}
\frac{d^2 \bm{r_i}}{dt^2} = -GM_{\rm U}\frac{\bm{r}_i}{|\bm{r}_i|^3}-\sum_{i \neq j} Gm_j\frac{\bm{r}_j-\bm{r}_i}{|\bm{r}_j-\bm{r}_i|^3}, 
\end{eqnarray}
where $\bm{r_i}$ and $m_i$ are respectively the position relative to the center of Uranus and mass of particle $i$.
We used a fourth-order Hermite scheme \citep{MA92} for time integration during the growth of inner particles, and also used the second-order Leap Frog method during the growth of outer particles. We adopted a shared time-step and changed it from $2^{-9}$$\Omega_{\rm R}^{-1}$ to $2^{-5}$$\Omega_{\rm R}^{-1}$ depending on particle growth.

Because the computational cost of calculating the gravitational interaction between all particles is expensive, we adopted the Framework for Developing Particle Simulator (FDPS), a library for particle-based numerical simulations, developed by \citet{Iwa16}. FDPS provides functions for efficient parallelization of calculations and reduces the calculation cost of the interaction from $\mathcal{O}(N^2)$ to $\mathcal{O}(N\log{N})$, where $N$ is the total number of particles introduced to the system..

The simulations consider interparticle collisions.
Specifically, such a collision is detected when the distance between two particles becomes smaller than or equal to the sum of their radii.
Collisions are assumed to be moderately inelastic.  The relative velocity of two colliding particles changes according to the following equation:
\begin{eqnarray}
 \bm{v_n}' &=& -\epsilon_n\bm{v_n} 
\label{eq:col1}\\
 \bm{v_t}' &=& \epsilon_t\bm{v_t} 
\label{eq:col2}
\end{eqnarray}
where $\bm{v}'$ and $\bm{v}$ are respectively the relative velocity after and before a collision and the subscripts $n$ and $t$ respectively represent normal and tangential components.
We set the normal component of the coefficient of restitution $\epsilon_n$ to 0.1 and the tangential component $\epsilon_t$ to 1, and we neglect particle spin for simplicity. The velocities of two particles after a collision are determined based on the law of conservation of momentum. 
Two particles must be separated as the distance between their centers equals the sum of their radii to avoid an unnecessary collision in the next time step; this separation is carried out under the law of conservation of angular momentum (see Appendix \ref{apdx:col}).

If the relative velocity of two particles after a collision is smaller than the surface escape velocity modified by the tidal force, they are gravitationally bounded.
In rotational coordinates around Uranus at distance $a$ with Kepler angular velocity $\Omega$, such conditions are described by negative Jacobi energy $E_{\rm J}$ of the two particles after the collision:
\begin{eqnarray}
E_{\rm J} = \frac{1}{2}|v|^2 -\frac{3}{2}x^2\Omega^2 + \frac{1}{2}z^2\Omega^2 - \frac{G(m_1+m_2)}{r} + \frac{9}{2}r_{\rm H}^2 \Omega^2 < 0,\label{eq:merge1}
\end{eqnarray}
where $m_1$ and $m_2$ are the masses of the two particles; $x$, $y$, and $z$ are the relative positions of the two particles; $r$ is the distance between the two particles, and $r_{\rm H}$ is the Hill radius defined by
\begin{eqnarray}
 r_{\rm H} = \left(\frac{m_1+m_2}{3M_{\rm U}}\right)^{\frac{1}{3}}a,
\end{eqnarray}
which is the region dominated by the attraction of the two particles.
The Jacobi energy can be negative even when the center of masses of the two particles is outside the Hill sphere.
For the two particles to be gravitationally bounded, the following condition must also hold:
\begin{eqnarray}
r_1 + r_2 \leq  r_{\rm H},\label{eq:merge2}
\end{eqnarray}
where $r_1$ and $r_2$ are the radii of the two particles. 
The two particles are gravitationally bounded when both conditions Eqs.(\ref{eq:merge1}) and (\ref{eq:merge2}) are satisfied \citep{Kok00}.

In the following $N$-body simulations, gravitationally bound particles are merged into one spherical particle.
The merging of the two particles is calculated based on the laws of conservation of total mass and momentum.
Collisional fragmentation of particles is not considered here because an increase in the number of particles greatly increases the calculation costs.

\subsection{Initial conditions of debris disks}

Just after a giant impact occurs, vaporized materials and the atmosphere of Uranus and the impactor are ejected, resulting in the presence of high-temperature gas disk as well as rock solids. 
The gas disk formed from the ejected materials around Uranus evolves through dynamical processes, chemical reactions, and radiative cooling.
Then the disk gas density would significantly decay due to the viscous diffusion before the ice condensation to form the debris disk.
We have investigated disk evolution after a giant impact to Uranus in detail \citep{Ida19}.

In this study, we focus on satellite accretion from debris disks of solids in a gas-free environment.
To investigate the types of debris disks suitable for in-situ formation of Uranian satellites, we simulated the evolution of debris disks with several initial conditions.
We considered the total mass and surface density distribution of debris disks as the most important factors in the satellite formation process.

We set the initial disk mass ($M_{\rm disk}$) to be several times the total mass of the current Uranian satellite system ($M_{\rm tot}\sim 10^{-4}M_{\rm U}$).
The surface density distribution is assumed to follow a power law with semimajor axis $a$ and is represented as $\Sigma(a) \propto a^{-q}$.
$q$ is set as 3.00, 2.15, 1.95, and 1.50 as inferred from the density profiles in the results by \citet{Keg18}.
Table \ref{tb:table1} shows the model sets of the initial disks with the masses and power-indexes of the surface density distribution.

\begin{table}[htb]
\centering
\caption{Model set of initial conditions}
  \begin{tabular}{ccr} \hline
    Model & $M_{\rm disk} [M_{\rm tot}]$ &  $q$ \\ \hline \hline
    Disk1 & 3  & 1.50  \\ 
    Disk2 & 4  & 2.15  \\ 
    Disk3 & 10 & 2.15  \\ 
    Disk4 & 4  & 1.95  \\ 
    Disk5 & 3  & 1.95  \\ 
    Disk6 & 3  & 3.00  \\ \hline
  \end{tabular}
\label{tb:table1}
\end{table}

The inner edge of the disk is Uranian radius. Although the outer edge of the disk is not shown explicitly in \citet{Keg18}, we set the outer edge to be 25$a_{\rm U}$ which includes the orbit of the outermost satellite, Oberon, by simply extrapolating from the results of \citet{Keg18}.
We assumed that the initial eccentricities and inclinations of disk particles follow a Rayleigh distribution.
The root-mean-square of the eccentricity $\langle e^2 \rangle^{\frac{1}{2}}$ is set to be 0.3 and that of the inclination $\langle i^2 \rangle^{\frac{1}{2}}$ is set to be 0.15.
The other orbital elements of disk particles are set randomly.
The number of disk particles is 10,000 in all models.
The density of disk particles is $\rho = 1.40$ g cm$^{-3}$ as inferred from the density of the Uranian satellite system.
Disk particles are assumed to be rigid spheres.
The physical radius of a disk particle is given by
\begin{eqnarray}
 r_{\rm{p}} = \left(\frac{m}{M_{\rm U}}\right)^{\frac{1}{3}}\left(\frac{\rho}{\rho_{\rm U}}\right)^{-\frac{1}{3}}\!\!\!\!R_{\rm U} = \frac{1}{2.456}\left(\frac{m}{M_{\rm U}}\right)^{\frac{1}{3}}\!\!\!\!a_{\rm R},
\end{eqnarray}
where $m$ is the particle mass.
The initial masses of particles are equal to each other.

\section{Results}\label{sc:result}
We performed $N$-body simulations of satellite formation for the six disk models.
The mass is normalized  by $M_{\rm tot}\sim10^{-4}M_{\rm U}$ and time is normalized by $T_{\rm K}$, which is the Kepler period at the distance of the Roche limit and is given by
\begin{eqnarray}
 T_{\rm K} = \frac{2\pi}{\Omega_{\rm R}} \approx 10.9 \mbox{  hour}.
\end{eqnarray}

\subsection{Mass distributions with semimajor axis}

\begin{figure}[htb]
  \centering
  \plotone{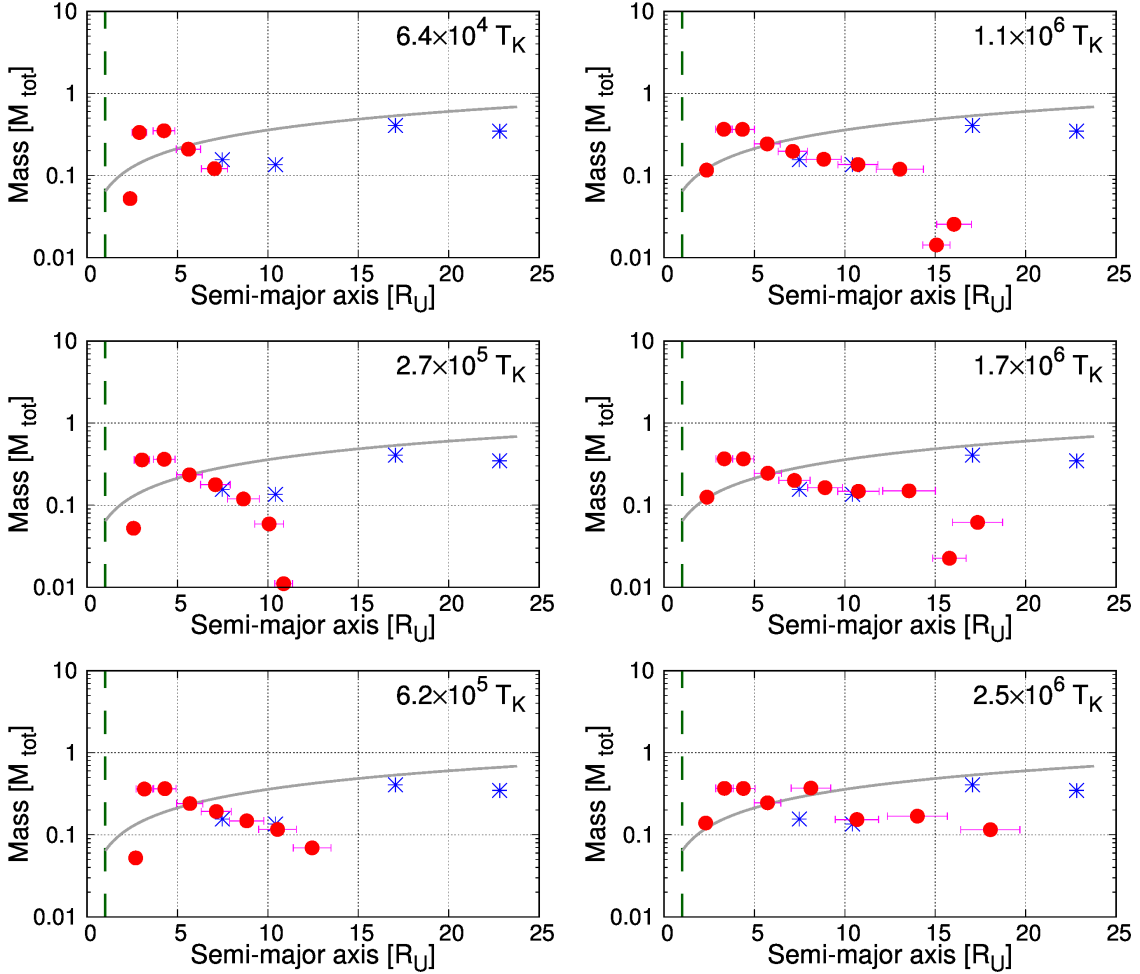}
  \caption{Time series of satellite mass distribution for Disk1 $[M_{\rm disk} = 3M_{\rm tot}, q = 1.50]$. The red filled circles represent grown-up particles in this simulation, and the lines from their centers to both sides have length of $5r_{\rm H}$. The blue stars represent Ariel, Umbriel, Titania, and Oberon in the current satellite system. The dashed  lines indicate the Uranian radius and the solid lines indicate the isolation mass distribution for the initial disk. 2.5$\times 10^6 T_{\rm K}$ equals $\sim$ 3100 years.}
\label{fg:disk1}
\end{figure}

Figure \ref{fg:disk1} shows the particle accumulation result for Disk1 [$M_{\rm disk}$ = $3M_{\rm tot}$, $q$ = 1.50], which is the mass distribution of grown-up particles and the outer four satellites (Ariel, Umbriel, Titania, and Oberon) differentiated by time from $ t = 6.4 \times 10^4 T_{\rm K} $ to $2.5 \times 10^6 T_{\rm K}$.
The isolation mass distribution for the initial disk is also shown in each panel (see Appendix \ref{apdx:iso}).
In the inner region of the disk, the local surface density and velocity dispersion of particles are larger than those in the outer region, and gravitational encounters between particles occur more often. Therefore, the particle growth timescale increases as the distance from Uranus increases. 

The grown-up particles at $t = 2.5 \times 10^6 T_{\rm K} $($\sim$ 3100 years) have comparable masses to each other, and their particle masses range from $0.1M_\mathrm{tot}$ to $0.4M_\mathrm{tot}$. 
With compared to the current satellites, the grown-up particles in the inner region 
 have much larger mass than the inner thirteen moons($<10^{-7}M_{\rm tot}$), and in the middle region ($3R_\mathrm{U}$ to $13R_\mathrm{U}$), several particles with a few times mass of Ariel or Umbriel are formed.
On the other hand, the two outermost grown-up particles have less masses and less orbital radii than the two outermost satellites, Titania and Oberon, respectively.
The total mass of the grown-up particles is around $1.9M_\mathrm{tot}$; this is around 64$\%$ of the initial disk mass. 
The mass that falls into Uranus from the disk is around $0.85M_\mathrm{tot}$; this is around 28 $\%$ of the initial disk mass. 

\begin{figure}[htb]
  \centering
  \plotone{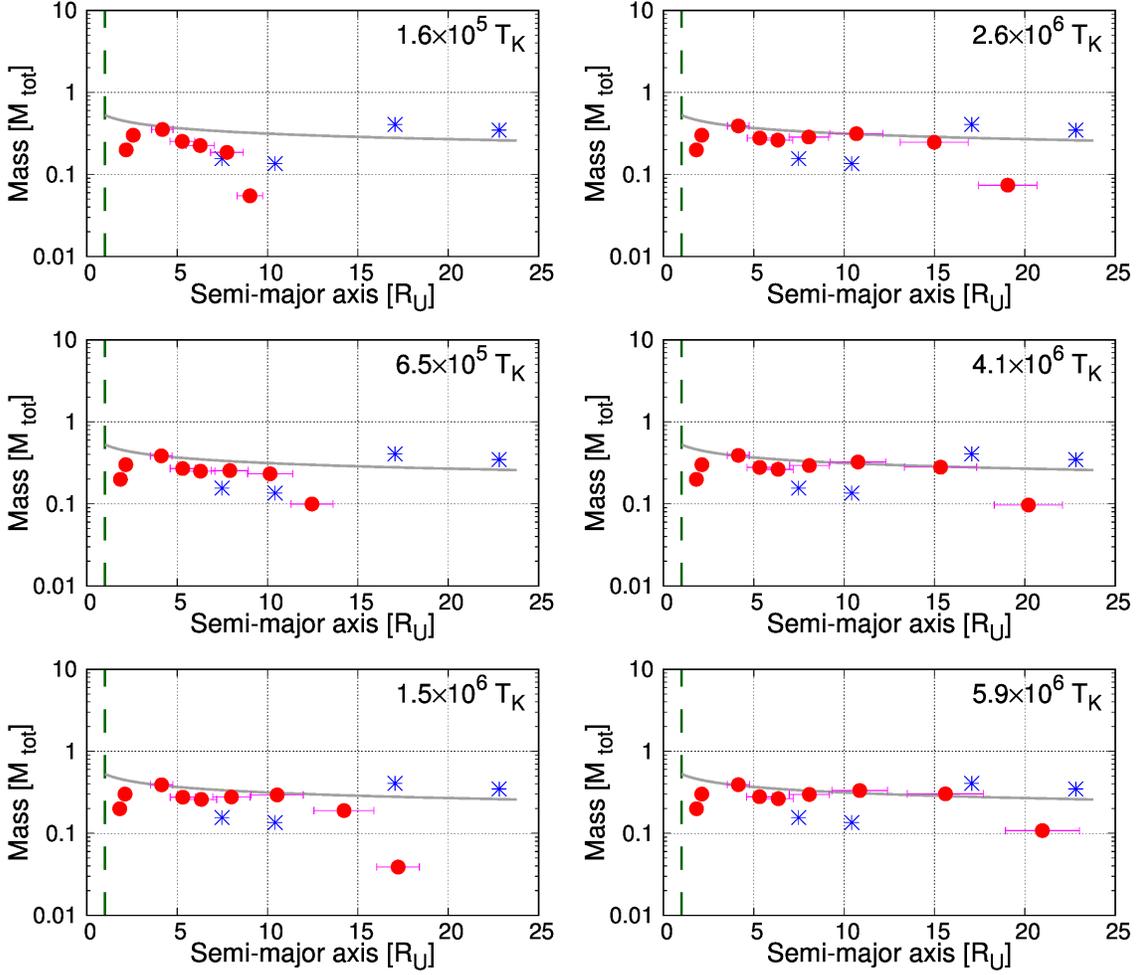}
  \caption{Same as Fig. \ref{fg:disk1} but for Disk2 [$M_{\rm disk}$ = $4M_{\rm tot}$, $q$ = 2.15].}
\label{fg:disk2}
\end{figure}

Similar results are obtained for Disk2 [$M_{\rm disk}$ = $4M_{\rm tot}$, $q$ = 2.15], shown in Fig. \ref{fg:disk2}, differentiated by time from $t = 1.6 \times 10^5 T_{\rm K}$ to $5.9 \times 10^6 T_{\rm K}$. 
Disk2 has slightly larger mass and larger value of $q$ than Disk1. Particles grow with timescales similar to those for Disk1.
The particle distribution is similar to that of Disk1; inner extra particles, more particles with higher masses in the middle, and two outermost particles with less masses and less orbital radii.
The total mass of the grown-up particles is around $2.4M_\mathrm{tot}$; this is around 61$\%$ of the initial disk mass. The mass that falls from the disk into Uranus is around $1.4M_\mathrm{tot}$; this is around 35 $\%$ of the initial disk mass.

\begin{figure}
\centering
\gridline{\fig{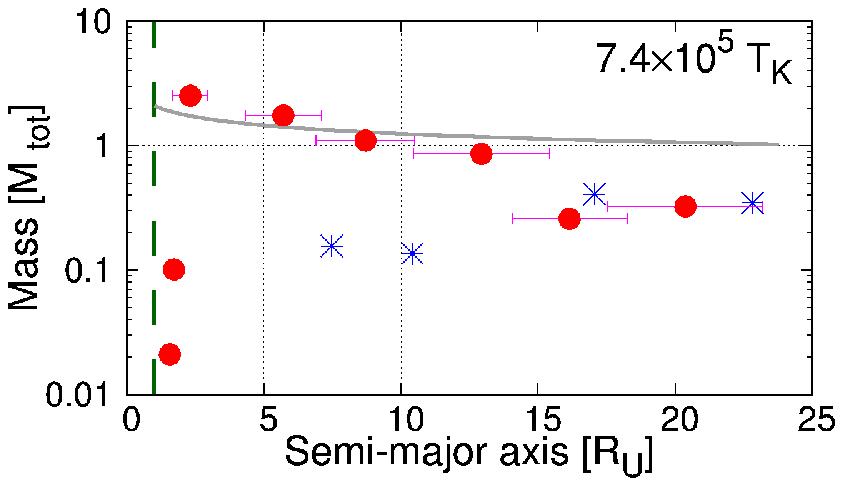}{0.45\textwidth}{(a) Disk3 [$M_{\rm disk}$ = $10M_{\rm tot}$, $q$ =  2.15]\label{fg:disk3}}
          \fig{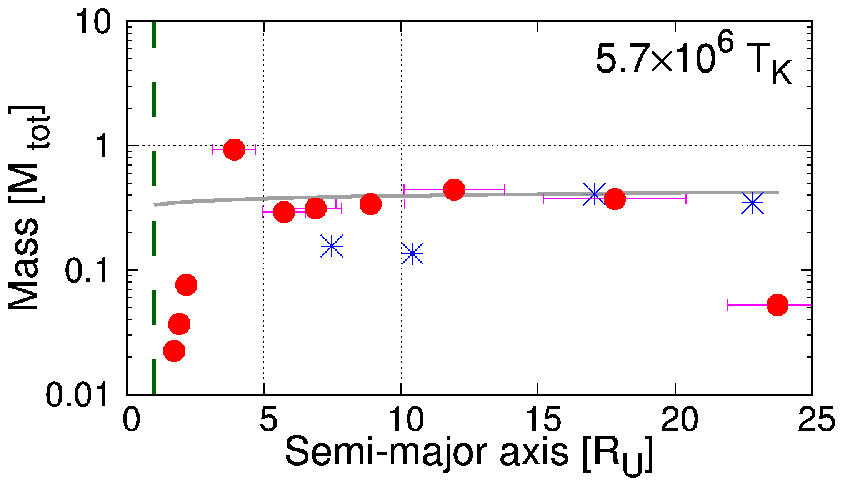}{0.45\textwidth}{(b) Disk4 [$M_{\rm disk}$ = $4M_{\rm tot}$, $q$ =  1.95]\label{fg:disk4}}}
\gridline{\fig{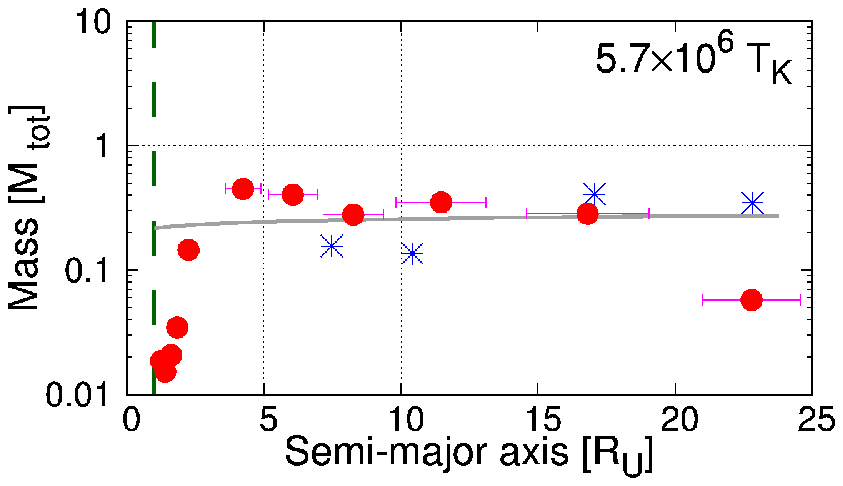}{0.45\textwidth}{(c) Disk5 [$M_{\rm disk}$ = $3M_{\rm tot}$, $q$ =  1.95]\label{fg:disk5}}
          \fig{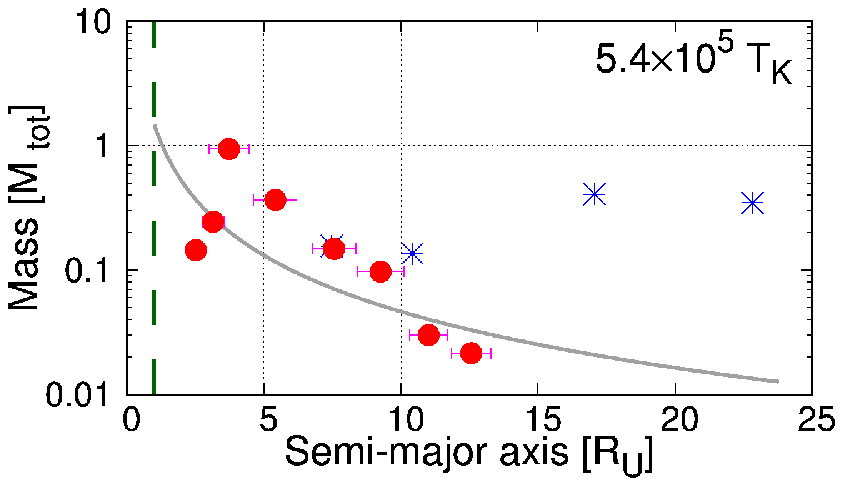}{0.45\textwidth}{(d) Disk6 [$M_{\rm disk}$ = $3M_{\rm tot}$, $q$ =  3.00]\label{fg:disk6}}}
\caption{Results for four disk models: (a) Disk3, (b) Disk4, (c) Disk5, and (d) Disk6.}\label{fg:disks}
\end{figure}

Figure \ref{fg:disks} shows the results for (a) Disk3 at $t = 7.4 \times 10^5 T_{\rm K}$, (b) Disk4 at $t = 5.7 \times 10^6 T_{\rm K}$, (c) Disk5 at $t = 5.7 \times 10^6 T_{\rm K}$, and (d) Disk6 at $t = 5.4 \times 10^5 T_{\rm K}$. 
Figure \ref{fg:disks}(a) shows that the total mass of the grown-up particles for Disk3 is $6.8M_\mathrm{tot}$; this is too much compared with that of the current satellites.
Figure \ref{fg:disks}(b) (Disk4) and Fig.\ref{fg:disks}(c) (Disk5) show that the initial conditions are similar but distributions are slightly different because of the stochastic effect during particle growth. 
In these results, the particle with comparable mass and orbit to Umbriel's formed, and the outermost particle has similar orbit of Oberon but still much less mass.
The sums of grown-up particles' mass in the outer region are $1.5M_\mathrm{tot}$ for Disk4 and $1.1M_\mathrm{tot}$ for Disk5.
Figure \ref{fg:disks}(d) (Disk6) shows that the mass of grown-up particles decreases sharply with the semimajor axis and that the mass distribution obviously differs from that of the current satellites.

Summarizing the results from the above-described disk models, the particle distribution from any above-described disk models could not directly reproduce the satellite distribution. However, the orbital evolution of particles with long timescale after their growth can alter the mass-orbit distribution. of the grown-up particles, so we take it into account in Section \ref{sc:dis}.

We note that particle growth in the above-described disk models has not been completed. Disk particles accounting for some percentage of the initial disk mass remain in each disk. 
Such particles can accrete to grown-up particles; alternatively, they can be cleared away from the system because of scattering with other particles in the later stage and can damp the eccentricities and inclinations of grown-up particles. 
However, they do not largely change the orbital radius and mass of grown-up particles.
Even if all remnant particles in orbits accrete to the outermost particle, in the all results here except for Disk3, its mass can not reach the mass of Oberon.

\subsection{Comparison with isolation mass}

When the power index $q$ is much larger than 2, as in the case of Disk6, the mass distribution shows a sharply decreasing slope with the semimajor axis and becomes greatly different from that of the current satellite system. Even if the effects of radial diffusion, orbital evolution, and some stochastic fluctuations are considered, the condition $q\geq 3$ is not suitable for in-situ formation of the current satellite system.

When the power index $q$ is around 2, as in the cases of Disk2, Disk3, Disk4, and Disk5, the mass distribution of the grown-up particles becomes almost flat. 
For example, for Disk2 (Fig.\ref{fg:disk2}), particles grow with the isolation mass distribution; however, there is a small difference owing to the effect of radial diffusion and stochastic fluctuation. 
The distributions for Disk4 and Disk5, especially in the inner regions, differ from each other despite the same power index because of the stochastic effect. 
The stochastic effect can arise largely from gravitational interactions in the packed orbits of the inner grown-up particles in the early stage; this is analogous to the giant impact regime in planetary formation. 

When the power index $q$ is less than 2, as in the case of Disk1 (Fig.\ref{fg:disk1}), the isolation mass is predicted to increase with the semimajor axis. 
However, the distribution of the isolation mass, especially in the inner region, greatly differs from that of grown-up particles. 
This may mainly be caused by mass transfer from the outer orbits due to radial diffusion. 

By using curve fitting analogous to least squares approximation, the isolation mass function (Eq.\ref{eq:iso}) can be fitted to the data distribution of the masses and the orbital radii of the outer four satellites. Then, $q$ and $M_{\rm disk}/M_{\rm tot}$ are fitted to be 1.36 and 1.76, respectively; these are close to the parameters for Disk1.
Even if $q <$ 2 including the fitted value for the outer four satellites, several extra satellites with larger masses than the isolation mass distribution could form in the inner region unless $q$ is a very low or negative value.

\section{Discussion}\label{sc:dis}

\subsection{Orbital evolution after satellite growth}

The particle distribution from the above-described disk models could not reproduce the mass-orbit distribution of the current satellites, but it can be affected by orbital evolution of particles after their growth.
Since the timescale of the orbital evolution is much larger than that of the satellite growth, which is less than thousands years, the main orbital evolutions would occur after the satellite growth. Orbital evolution of a satellite mainly occurs owing to gas drag in satellite orbits, tidal torque from a central planet, tidal dissipation in the interior of the satellite, and gravitational interaction with other satellites.

Our calculations do not consider the existence of gas. 
Gas infall from the interplanetary region can be considered a possible cause of the existence of gas in the circum-Uranus region. 
However, in the giant impact regime of the formation of the Solar system, gas may have dissipated from the Uranian orbit (e.g., \citealp{LS93}). 

A tidal torque is caused by a difference between the rotational period of Uranus and the orbital period of a satellite. 
Angular momentum is transferred between the planet and the satellite, and therefore, the satellite orbit evolves. 
The orbital radius of a satellite whose orbital period corresponds to the rotational period of Uranus $T_{\rm U}$ is called the corotation radius $r_{\rm c}$, and it is expressed as
\begin{eqnarray}
r_{\rm c} = \left(\frac{T_{\rm U}}{2\pi}\right)^\frac{2}{3} (GM_{\rm U})^\frac{1}{3}.
\end{eqnarray}
At present, $T_{\rm U} = $17h 14m 24s = 62064s; then, $r_{\rm c}$ = 3.30 $R_{\rm U}$. 
The corotation radius normalized by the Roche limit is $r_{\rm c}$ = 1.39 $a_{\rm R}$ for $\rho = 1.40$.
A satellite inside the corotation radius receives negative torque from Uranus and migrates inward, whereas  one outside the corotation radius receives positive torque and migrates outward.

Tidal dissipation occurs in the interior of a satellite with an eccentric orbit owing to the tidal force generated by a planet; it can act to heat the satellite and damp its eccentricity. 
The tidal torque depends on the composition of Uranus, and tidal dissipation mainly depends on the composition of the satellite.

The semimajor axis $a$ and the eccentricity $e$ of a satellite evolve according to the following equations \citep{Cha10}:
\begin{eqnarray}
\frac{\mathrm{d} a}{\mathrm{d}t} &=& \mathrm{sgn}(a-r_{\mathrm{c}}) \frac{3k_{2\mathrm{p}}MG^{1/2}R_{\mathrm{p}}^5}{Q_{\mathrm p}M_{\mathrm{p}}^{1/2}a^{11/2}}\left(1+\frac{51e^2}{4}\right)
-\frac{21k_{2}n M_{\mathrm{p}}R^5}{Q M a^{4}}e^2,\label{eq:-a}\\
\frac{\mathrm{d} e}{\mathrm{d}t} &=& \mathrm{sgn}(a-r_{\mathrm{c}}) \frac{57k_{2\mathrm{p}}nMR_{\mathrm{p}}^5}{8Q_{\mathrm p}M_{\mathrm{p}}a^{5}}e
-\frac{21k_{2}n M_{\mathrm{p}}R^5}{2Q M a^{5}}e,\label{eq:-e}
\end{eqnarray}
where $k_{2\mathrm{p}}$ ($k_{2}$), $Q_{\mathrm p}$ ($Q$), $M_{\mathrm{p}}$ ($M$), and $R_{\mathrm{p}}$ ($R$) are the tidal Love number, tidal quality factor, mass of the planet, and radius of the planet  (satellite), respectively, and $n$ is the satellite's orbital frequency. 
The first term in each equation accounts for tidal torque from Uranus and the second term accounts for tidal dissipation in the interior of the satellite. 
For a satellite orbiting inside (outside) the corotation radius, the first term becomes negative (positive).
These evolution rates largely depend on the satellites' semimajor axes.

\begin{figure}[htb]
  \centering
  \plottwo{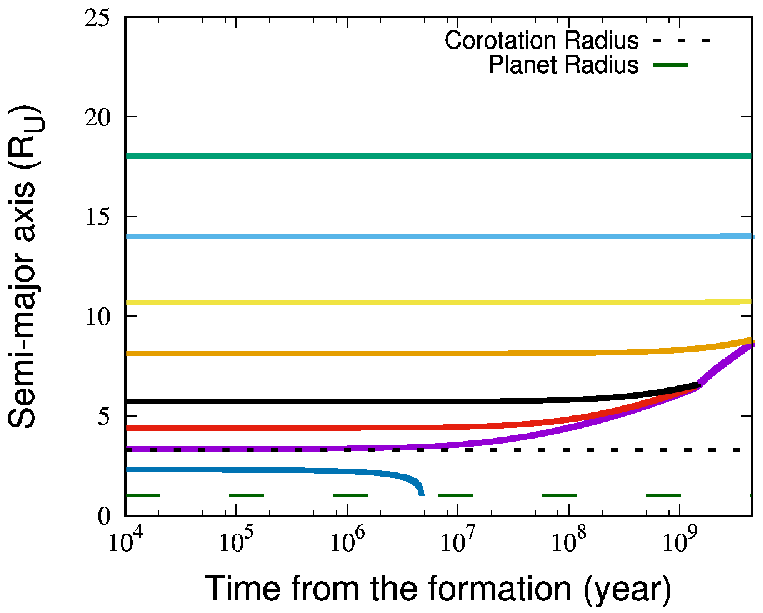}{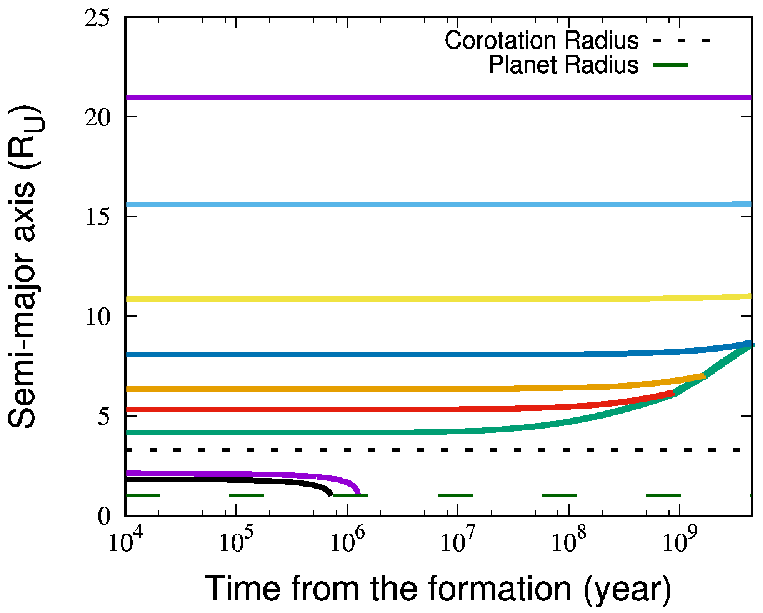}
  \caption{Tidal evolution of orbits of grown-up particles for Disk1 (left panel) and Disk2 (right panel). Solid lines indicate orbital evolution of particles with time, and long and short dashed lines respectively indicate the Uranian radius and corotation radius. Here, gravitational interaction between particles is not considered. When particles' orbits overlap, they are merged into one based on the law of mass conservation.}\label{fg:orbit}
\end{figure}

We analytically calculated the orbital evolution of grown-up particles over 4.5 billion years for Disk1 and Disk2 according to Eqs.(\ref{eq:-a}) and (\ref{eq:-e}); the results are shown in Fig.\ref{fg:orbit}.
We set the tidal Love number of Uranus as $k_{2\mathrm{p}} = 0.104$ \citep{GZ77}; tidal quality factor of Uranus as $Q_{\mathrm p} = 11,000$, which is the lower limit of the constrained value by \citet{TW89}; and tidal parameters of the particles as $k_{2}/Q = 10^{-5}$ with reference to \citet{TW89}.
In these analytic calculations, the gravitational interaction between particles is not considered and the corotation radius is assumed to be the same as the present one during orbital evolution.

As shown in Fig.\ref{fg:orbit}, particles inside the corotation radius fall to the planet over several million years whereas particles in the region $r_{\mathrm c} \leq a \lesssim 10R_{\mathrm U}$ move outward and merge each other in several billion years. By contrast, particles in the region $a \gtrsim 10R_{\mathrm U}$ would mostly remain in their orbits over billions of years. This is because the evolution rate of a particle's orbit is affected largely by its semi-major axis rather than by its mass. According to the orbital evolution of satellites owing to tides, inner extra particles around the corotation radius would move out of this region. 
However, the particles in the middle region would increase its mass to several times after migration and merger, so it would eventually obtain further mass compared to Ariel or Titania.
On the other hand, the outermost satellites would not change their orbits and masses.

Inside the corotation radius, the particles which migrate inward would be disrupted by Uranus' tidal force, and their fragments would form a ring inside the Roche limit.
The inner small satellites can form from such rings based on the model proposed by \citet{CC12}.
According to this model, in order to form these inner small satellites, the mass of the satellite-forming ring would need $\sim 1.06 \times 10^{-3} M_{\rm tot}$.

\citet{HM19} suggested that, under the ring-satellite system, Miranda could have formed and evolved out only to about 4 Uranian radii, and not to 5 Uranian radii where it is now (as ring torques cannot act beyond the 2:1 outer resonance with ring edge at the Roche limit). So Miranda may have originally formed close to 5 Uranian radii where it is now.

Conclusively we suggest that, even if the effect of the orbital evolution is taken into account, the orbital distribution of the five major satellites could not be reproduced from the above-described disk, where the power index of its surface density is similar to that of the disk generated just after the giant impact.
In order to explain that outer two satellites (Titania and Oberon) exceed the two middle satellites (Ariel and Umbriel) in their masses through in-situ formation, the initial debris disk should have a mass distribution of solids that increases with distance from Uranus, which seems counter-intuitive for material distribution inferred from giant impact simulations.
We speculate that a evaporated disk after a giant impact would experience some thermal and viscous evolution, and then the five massive satellites would form from a disk of solids whose $q$ value is much less than 1.5 or negative (namely, the power index of its surface density is positive), whose outer edge reach around the orbit of Oberon.
Even in such a situation, orbital migration of satellites inside the corotation radius can occur, so if a satellite inside the corotation radius have $10^{-3} M_{\rm tot}$ at least, it can migrate into the Roche limit within 4.5 billion years, and then the small inner satellites are still possible to form from rings generated by the disruption of it.

\section{Conclusion}

We modeled a wide debris disk generated by the giant impact, performed $N$-body simulations of satellites accretion in such a disk, and investigated the possibility of the in-situ formation of the Uranian satellites, taking account of the orbital evolution of satellites due to the planetary tides after their growth.
We found that, from such disks, satellites in the middle region ($3R_\mathrm{U}$ to 13$R_\mathrm{U}$) would have much larger masses than Ariel or Umbriel, and the outermost satellite would not obtain a mass of Oberon, so the orbital distribution of the five major satellites could not be reproduced.

However, we still speculate that the five major satellites would form in the current site since it would be difficult to form from rings inside the Roche limit and migrate to the current orbits. We also speculate that a evaporated disk after a giant impact would experience some thermal and viscous evolution, and then the five massive satellites would form from a disk whose $q$ value is much less than 1.5 or negative, whose outer edge reach around the orbit of Oberon. On the other hand, the small inner satellites may form from rings generated by the satellites which moved inward and disrupted by the planetary tides \citep{CC12}.

In future studies, it would be necessary to investigate the thermal and viscous evolution of a evaporated disk generated just after an impact into an icy giant, and then simulate the satellite formation with $q$ be much smaller and negative to realize the in-situ formation of the Uranian satellites.

\acknowledgments

We thank the anonymous reviewer's constructive comments, which led us to greatly improve this paper. The $N$-body simulation in this study was conducted at the Yukawa Institute Computer Facility. This work was supported by JSPS KAKENHI grant number 19K03950 and KURA Research Development Program ISHIZUE.

\appendix

\section{Separation between two colliding particles}\label{apdx:col}
When the distance between two particles is smaller than or equal to the sum of their radii, the collision is detected.
If the distance between two particles is still smaller than the sum of their radii in the next time step, an unnecessary additional collision can be detected by mistake.
In order to avoid such unnecessary collision detections, two particles should be separated for the distance between them to be equal to the sum of their radii on the basis of conservation of angular momentum.

\begin{figure}[htb]
  \centering
  \plotone{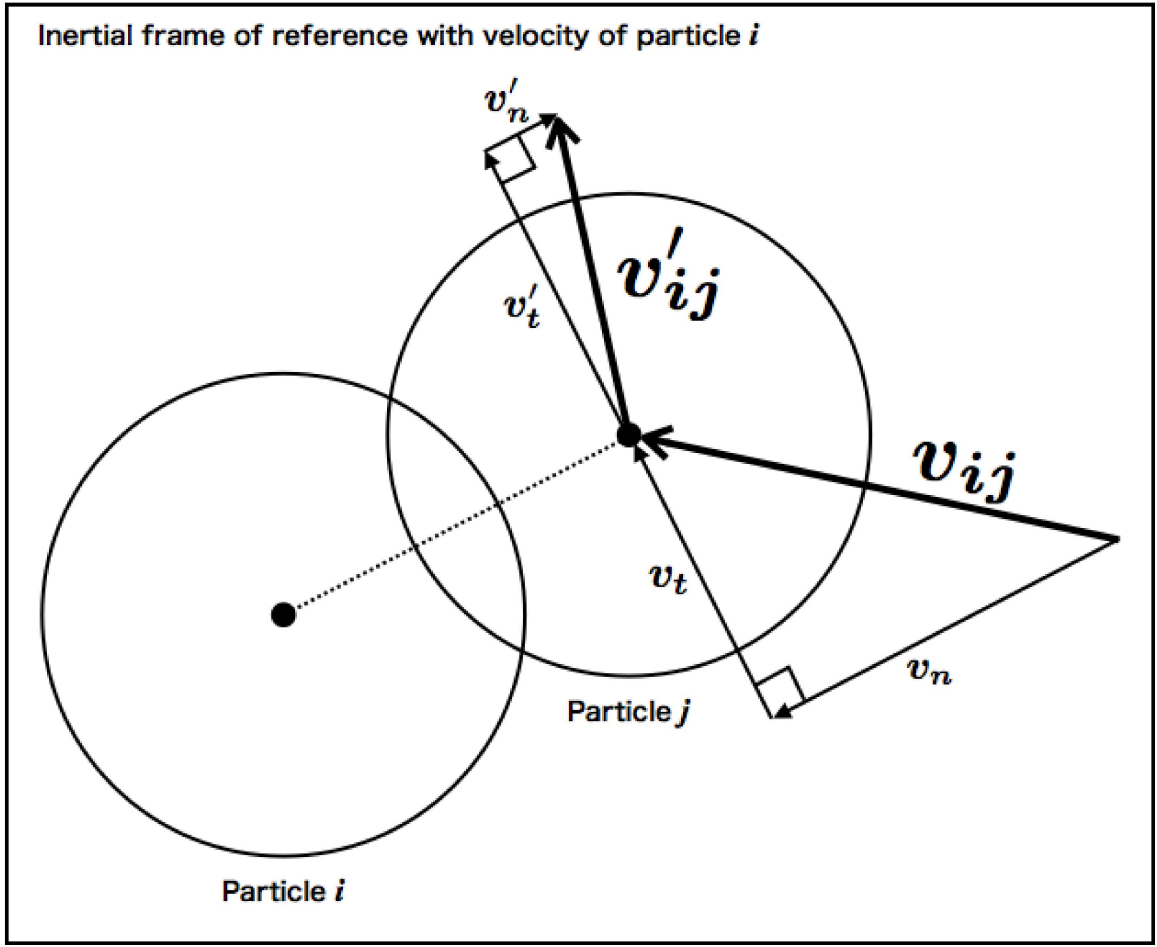}
  \caption{A sketch of a rebound between two colliding particles in the inertial frame of reference with the impact velocity of particle $i$. }
\label{fg:colv}
\end{figure}

Hereafter two colliding particles, particle $i$ and particle $j$, are considered.
First, after two particles collide with each other, their velocities are changed based on Eqs.(\ref{eq:col1}), (\ref{eq:col2}) and conservation of momentum as given by
\begin{eqnarray}
 m_i\bm{v_i}' + m_j\bm{v_j}' =  m_i\bm{v_i} + m_j\bm{v_j},
\end{eqnarray}
where $m$, $\bm{v}$ and $\bm{v}'$ are mass, impact velocity, and rebound velocity of a particle, respectively. The subscripts represent a kind of particles, $i$ or $j$.
Figure \ref{fg:colv} shows a sketch of a collision, where $\bm{v_{ij}} = \bm{v_j} - \bm{v_i}$ and $\bm{v_{ij}'} = \bm{v_j}' - \bm{v_i}'$.

\begin{figure}[htb]
  \centering
  \plotone{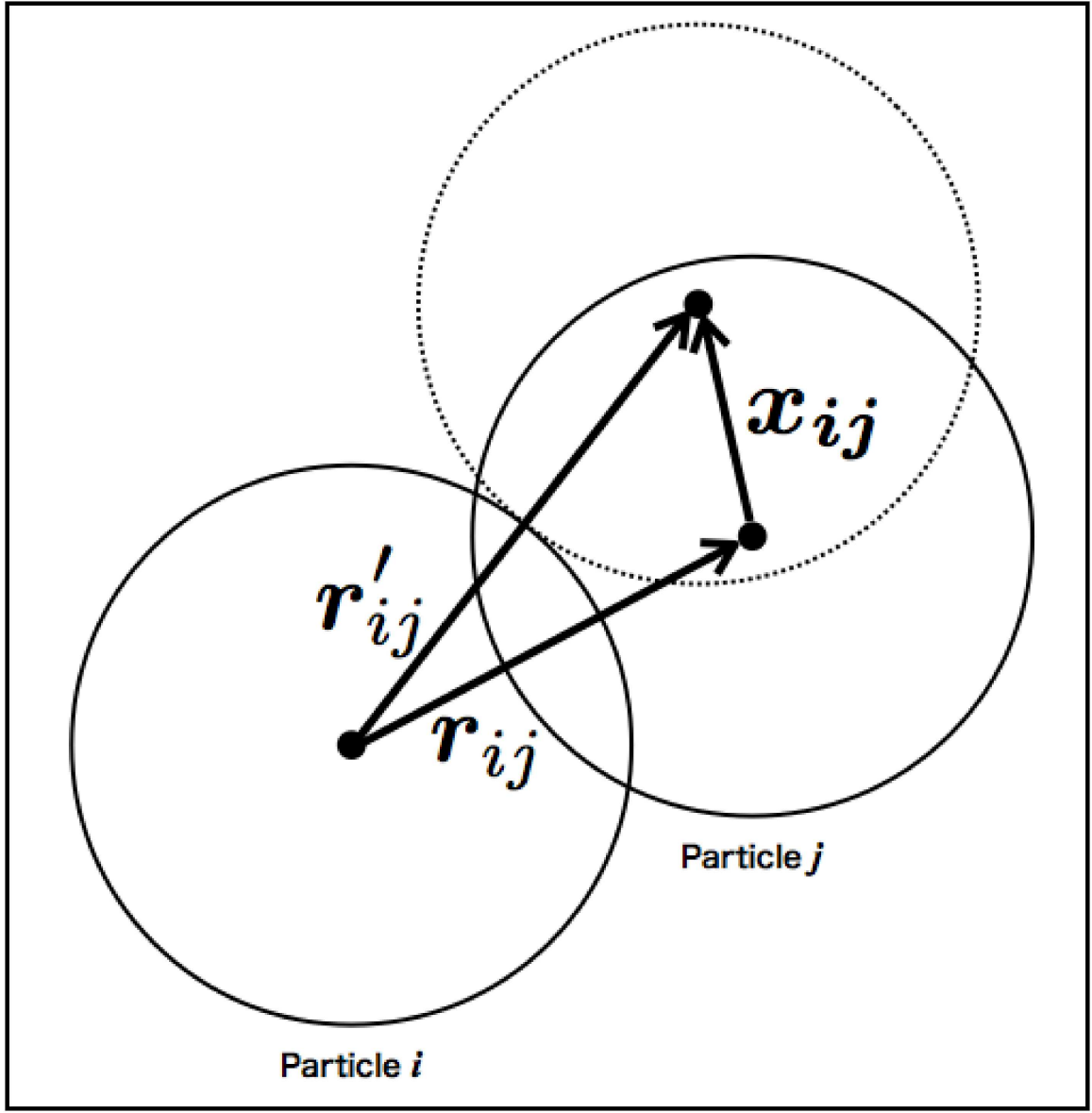}
  \caption{A sketch of a separating two particles.}
\label{fg:colr}
\end{figure}

Second, a separation between of two particles after a collision is carried out based on conservation of angular momentum,
\begin{eqnarray}
 m_i \bm{r_i}' \times \bm{v_i}' + m_j\bm{r_j}' \times\bm{v_j}'=m_i\bm{r_i} \times\bm{v_i}' + m_j\bm{r_j} \times\bm{v_j}',
\end{eqnarray}
where $\bm{r}$ and $\bm{r}'$ indicate orbital radii of particles before and after a separation, respectively, and also based on the following equations;
\begin{eqnarray}
\bm{r_{ij}'} &=& \bm{r_{ij}} + \bm{x_{ij}},\\
\bm{x_{ij}} & \parallel & \bm{v_{ij}'},
\end{eqnarray}
where $\bm{r_{ij}} = \bm{r_j} - \bm{r_i}$, $\bm{r_{ij}'} = \bm{r_j}' - \bm{r_i}'$, and $\bm{x_{ij}}$ is a modifying vector as represented in Fig.\ref{fg:colr}, which is set to be parallel to the relative rebound velocity.

Then $\bm{x_{ij}}$ is determined by,
\begin{eqnarray}
\bm{x_{ij}} &=& \left( -\frac{\bm{r_{ij}}\cdot\bm{v_{ij}}}{|\bm v_{ij}|} +\sqrt{\left(\frac{\bm{ r}_{ij}\cdot\bm{v_{ij}}}{|\bm v_{ij}|}\right)^2 + r_{{\rm p}ij}^2 - |\bm{r_{ij}}|^2}\right) \frac{\bm{v_{ij}}}{|\bm v_{ij}|},
\end{eqnarray}
where $r_{{\rm p}ij}$ is the sum of the particles' radii.

Finally, each modified orbital radius $\bm{r}'$ is given by,
\begin{eqnarray}
\alpha_{ij} &=& \frac{m_i(\bm{x_{ij}}\times \bm{v_j}')\cdot(\bm{x_{ij}}\times \bm{v_i}')}{m_j|\bm{x_{ij}}\times \bm{v_j}'|^2},\\
 \bm{r_i}' &=& \bm{r_i} -\frac{1}{\alpha_{ij} + 1}\bm{x_{ij}},\\
 \bm{r_j}' &=& \bm{r_j} +\frac{\alpha_{ij}}{\alpha_{ij} + 1}\bm{x_{ij}}.
\end{eqnarray}

\section{Isolation Mass}\label{apdx:iso}

The isolation mass, $M_\mathrm{iso}$, is the asymptotic mass derived from the basic analysis model when neglecting the radial diffusion and orbital evolution of the satellites. It is expressed by the following equation with reference to the core accretion model of planetary formation (e.g. \citealt{Lis87,KI98,KI00}):
\begin{eqnarray}
M_\mathrm{iso} &=& 2\pi a \cdot 10r_{\rm H} \cdot \Sigma,
\end{eqnarray}
where $\Sigma$ is the surface density of satellitesimals, which is given by
\begin{eqnarray}
\Sigma = \cfrac{M_{\rm disk}}{\int^{a_{\rm max}}_{a_{\rm min}} 2\pi a^{1-q}\mathrm{d}a}a^{-q},
\end{eqnarray}
and $r_{\rm H}$ is the Hill radius of an isolation mass, which is given by $(2M_{\rm iso}/3M_{\rm U})^{1/3}$.
Therefore,
\begin{eqnarray}
M_\mathrm{iso} \simeq
\left\{
\begin{array}{l}
0.26\times \left[\cfrac{2-q}{a_{\rm max}^{2-q} - a_{\rm min}^{2-q}}\right]^{\frac{3}{2}}\!\!\!\!\left(\cfrac{M_{\rm disk}}{M_{\rm tot}}\right)^{\frac{3}{2}}\!\!\left(\cfrac{a}{a_{\rm R}}\right)^{\frac{3}{2}(2-q)}\!\!\!\!\!\!\!\!\!\!M_{\rm tot} \:\:\:\: \mbox{($q$ $\neq$ 2)}\\
0.26\times \left[\ln\left(\cfrac{a_{\rm max}}{a_{\rm min}}\right)\right]^{-\frac{3}{2}} \left(\cfrac{M_{\rm disk}}{M_{\rm tot}}\right)^{\frac{3}{2}}\!M_{\rm tot} \:\:\:\: \mbox{($q$ = 2)} 
\end{array},
\right.\label{eq:iso}
\end{eqnarray}
where $a_{\rm max}$ and $a_{\rm min}$ are the semimajor axes of the outer and the inner edge of a satellite-forming disk, respectively.
Eq.(\ref{eq:iso}) indicates that if $q>2$, the isolation mass increases with the semimajor axis, whereas if $q<2$, the isolation mass decreases with the semimajor axis. 
If $q=2$, the isolation mass does not depend on the semimajor axis.

\end{document}